\documentclass[12pt]{article}
\usepackage{amsmath,a4}
\usepackage{amsfonts,amssymb}
\usepackage{amscd}
\usepackage{eucal,mathrsfs}
\usepackage[hypertex]{hyperref}
\usepackage{amsthm}
\usepackage{graphics}
\usepackage{graphicx}
\usepackage{pstricks}
\numberwithin{equation}{section}

\newtheorem{theorem}{\rmfamily\bfseries{Teorema}}[section] 
\newtheorem{corollary}[theorem]{\rmfamily\bfseries{Corollary}}

\newtheorem{proposition}[theorem]{\rmfamily\bfseries{Proposition}}

\def\un{1\kern-3pt \rm I}
\allowdisplaybreaks[4]
\date{}
\makeatother
\makeatletter
\@addtoreset{equation}{section}
\advance\hoffset by -7mm   
\makeatletter 
\@addtoreset{equation}{section}  
\makeatother

\def\ptoday{{\ifcase\month 
\or January, \or February, \or March, \or April,\or May, 
\or June, \or July, \or August, \or September, \or October, 
\or November, \or December,\fi\ \number \year}}


\begin{document}

%%%%%%%%%%%%%%%%%%%%%%%%%%%%%%%%%%%%%%%%%%%%%%%
\title{Moduli Structures, Separability of the Kinematic Hilbert Space 
and Frames in Loop Quantum Gravity}

\author{Bruno Carvalho %\footnote{Work supported by the
%Coordenação de Aperfeiçoamento de Pessoal de Nível Superior --
%CAPES (Brazil)}
 and Daniel H.T. Franco%\footnote{Work partially funded by the 
%Fundação de Amparo à Pesquisa do Estado de Minas Gerais -- 
%FAPEMIG (Brazil) and 
%the Conselho Nacional de Desenvolvimento Cient\'{\i}fico e
%   Tecnol\'{o}gico -- CNPq (Brazil)}
\\[4mm]
{\small Departamento de Fisica, 
Universidade Federal de Vi\c cosa -- UFV}\\{\small 
 Vi\c cosa, MG, Brazil}}
   
%\date{February 2015}
\date{\today}

\maketitle

\begin{center} 

\vspace{-5mm}

{\small{ e-mails: bruno.lqg@gmail.com, daniel.franco@ufv.br} }
\end{center}

%%%%%%%%%%%%%%%%%%%%%%%%%%%%%%%%%%%%%%%%%%%%%%%%%%%%%%%%%%%%%%%%%%%%%%%%%%%%%%%%
%\title{Moduli structures, separability of the kinematic Hilbert space 
%and frames in loop quantum gravity}
%\titlerunning{Kinematic Hilbert space 
%and frames in loop quantum gravity}

%\author{Daniel H.T. Franco \and Bruno C. Neves}

%\institute{Universidade Federal de Vi\c cosa, Departamento de F\'\i sica,\\
 %          Avenida Peter Henry Rolfs s/n, Campus Universit\'ario,\\
  %         Vi\c cosa, MG, Brasil, CEP:36570-900.
%\\
%\email{daniel.franco@ufv.br, bruno.lqg@gmail.com}}

%\authorrunning{Bruno C. Neves and Daniel H.T. Franco}

%\date{Received:  / Accepted: }

%\communicated{}

\begin{abstract}
We reassess the problem of separability of the kinematic Hilbert 
space in loop quantum gravity under a new mathematical point of view. We use the 
formalism of frames, a tool used in signal analysis, in order to remove the redundancy 
of the moduli structures in high valence graphs, without resorting to set extension 
of diffeomorphism group. For this, we introduce a {\it local} redundancy which encodes
the concentration of frame vectors on the tangent spaces $T_pM$ around points of 
intersections $p$ of smooth loops $\alpha$ in $\mathbb{R}^{3}$.
\end{abstract}

%\keywords{Moduli structures, separability of the Hilbert space, frames, loop quantum gravity}
%\\
%{\bf Mathematics Subject Classification (2010):} 42C40, 32A45, 46F15, 46F20.

%\newpage

%%%%%%%%%%%%%%%%%%%%%%%%%%%%%%%%%%%%%%%%%%%%%%%%%%%%%%%%%%%%%%%%
\section{\bf Introduction}
\label{Sec1}
%\hspace*{\parindent}
\parindent3em
%%%%%%%%%%%%%%%%%%%%%%%%%%%%%%%%%%%%%%%%%%%%%%%%%%%%%%%%%%%%%%%%
Quantum gravity is expected to reconcile general relativity and quantum
mechanics. Loop quantum gravity (LQG)~\cite{Rov,Th} is a proposal for a 
non-perturbative and background independent quantization of general relativity. 
In LQG the quantization of space is described in terms of a basis of states known 
as spin-network states denoted by $s$-knots. They are labeled by quantum numbers 
which ensures a bijection with the Natural numbers. Nevertheless, this bijection, 
which characterizes the se\-pa\-rability of the set of states, can be violated. 
In fact, a few years ago, Fairbairn and Rovelli~\cite{FR} studied the problem of the 
separability of the background-independent space of the quantum states of the 
gravitational field (the kinematical Hilbert space ${\mathscr H}_{\rm diff}$) in 
LQG. They have shown that ${\mathscr H}_{\rm diff}$ is {\it nonseparable} and
redundant (physically meaningless -- moduli) as a consequence of the fact that the 
knot space of the equivalence classes of graphs 
under diffeomorphisms are not countable. To circumvent this problem and
get rid of the redundant parameters, they proposed
a small extension of the group of homeomorphisms that are smooth (with their 
inverse) except possibly at a finite number of points. The space of the knot
classes become countable and ${\mathscr H}_{\rm diff}$ is separable.
In this contribution, we propose to (re)-examine the question of separability of 
${\mathscr H}_{\rm diff}$ using another mathematical setting. We shall use the 
formalism of frames, a tool used in signal analysis, in order to remove the redundancy 
of the moduli structures in high valence graphs, without resorting to set extension of 
diffeomorphism.

%%%%%%%%%%%%%%%%%%%%%%%%%%%%%%%%%%%%%%%%%%%%%%%%%%%%%%%%%%%%%%%%
\section{\bf Moduli structure of knots with intersections:
the Grot-Rovelli case}
\label{Sec2}
\parindent3em
%\hspace*{\parindent}
%%%%%%%%%%%%%%%%%%%%%%%%%%%%%%%%%%%%%%%%%%%%%%%%%%%%%%%%%%%%%%%%
Many of the most important problems in mathematics concern classification. 
One has a class of mathematical objects and a notion of when two objects
should count as equivalent. It may well be that two equivalent objects look superficially 
very different, so one wishes to describe them in such a way that equivalent objects 
have the same description and inequivalent objects have different descriptions.
In this section, we shall illustrate some of the key features of moduli structure of 
knots with intersections in LQG.

\begin{figure}[!htb]
 \centering
 \includegraphics[scale=0.40]{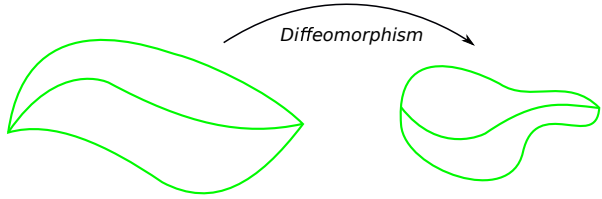}
 \caption{Schematic illustration of a diffeomorphism equivalence class.}
 \label{f.1}
\end{figure}

To begin with, we recall that in LQG the quantum states are labeled by knots with intersections~\cite{Rov}. 
These knots can be defined in two distinct ways: either as equivalence classes 
in $\mathbb{R}^{3}$ under a continuous deformations of the loop image --
``$c$-knots'' -- or as equivalence classes under diffeomorphisms of $\mathbb{R}^{3}$  --
``$d$-knots'' (see Figure \ref{f.1}). 
However, Grot-Rovelli showed in Ref.~\cite{GR} that such knots equivalence 
classes of loops in $\mathbb{R}^{3}$ under diffeomorphisms are not countable; rather, they 
exhibit a moduli-space structure. So, the two definitions 
lose their equivalence when the knots have intersections, and $d$-knots
are different from $c$-knots such that the space $\mathscr{K}_{d}$ of $d$-knots
is not countable. The continuous structure of the $\mathscr{K}_{d}$ comes from the differential 
structure of the underlying manifold, which means the presence of tangent spaces $T_pM$ at
intersection points. The loops define lines in $T_pM$ and the diffeomorphisms act linearly 
on $T_pM$. Hence, diffeomorphism equivalence implies equivalence under linear transformations 
of $T_pM$. Therefore, for a large enough number of lines the linearity is lost, 
since we can not align all lines by a linear transformation if the set of parameters 
is greater than the dimension of the vector space considered.

\begin{figure}[!htb]
 \centering
 \includegraphics[scale=0.40]{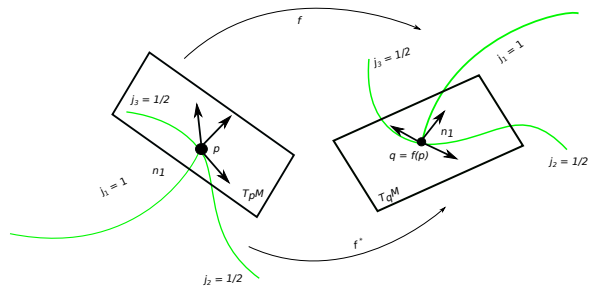}
 \caption{Equivalence of diffeomorphisms and linear transformations between
 $T_pM$ and $T_qM$.}
 \label{f.2}
\end{figure}

\begin{figure}[!htb]
 \centering
 \includegraphics[scale=0.30]{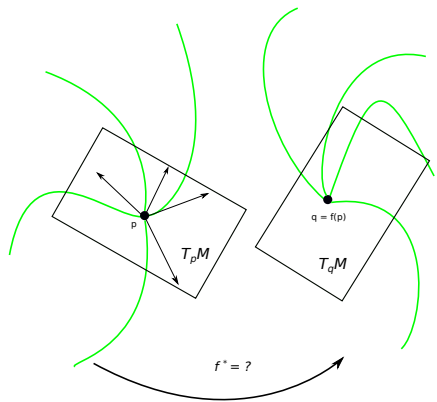}
 \caption{Schematic illustration of the diffeomorphism problem in high valence graphs.}
 \label{f.3}
\end{figure}

We will illustrate in more detail how this comes, by considering the example of 
Grot-Rovelli~\cite{GR}. Consider a smooth loop $\alpha$ in $\mathbb{R}^{3}$, with a 
point of self-intersection $p$ $\in$ $\mathbb{R}^{3}$ 
and assume that $\alpha$ passes five times through $p$, such that, at this point 
we have five tangent vectors 
$\boldsymbol{v}_{1},\ldots,\boldsymbol{v}_{5}$. Let us consider three of them 
linearly independent and 
denote by $\mathscr{K}_{c}[\alpha]$ the $c$ knot to which $\alpha$ belongs. 
Let now $\beta$ be an element in the same $\mathscr{K}_{c}[\alpha]$ and, as $\alpha$, 
it shall have five tangent vectors denoted by $\boldsymbol{w}_{1},\ldots,\boldsymbol{w}_{5}$ 
in a point $q$. In order to guarantee 
that $\alpha$ and $\beta$ are in the same $d$-knot, there must be a diffeomorphism 
$f:\mathbb{R}^{3} \rightarrow \mathbb{R}^{3}$ which transforms $\alpha$ into $\beta$. The 
tangent application $f^{\ast}$ maps vectors of the tangent space at $p$, $T_pM$, into 
tangent space at $q$, $T_qM$, and it should align the tangent vectors $\boldsymbol{v}_{i}$ 
to the tangent vectors $\boldsymbol{w}_{i}$ (see Figure \ref{f.2}). 
However, $f^{\ast}$ is a linear application between three-dimensional tangent spaces, given by 
the Jacobian of the matrix of $f$ at $p$ which depends on nine parameters. Since the directions 
of the five vectors $\boldsymbol{v}_{i}$ depend on ten parameters, it is easily seen that no linear 
transformation exists that aligns such five given vectors $\boldsymbol{v}_{i}$ to five 
given vectors $\boldsymbol{w}_{i}$ (see Figure \ref{f.3}). 
Generally, $\alpha$ and $\beta$ shall not belong 
to the same $d$ knot. Hence, there exists a continuous parameter which is function of 
the angle among the five tangents - which is invariant under diffeomorphism and 
distinguishes $\alpha$ from $\beta$. This is the root of {\it moduli} spaces. 
In other words, we see that our problem is related to the fact that there are linearly 
dependent vectors which makes the problem overcomplete, which means that the 
set becomes {\it redundant}. This redundancy is responsible for the lost of separability of 
${\mathscr H}_{\rm diff}$. In order to restore the separability and to avoid singular points 
Fairbairn and Rovelli~\cite{FR} proposed a small extension of the group of homeomorphisms 
that are smooth, with their inverse, except possibly at a finite number of points. In what follows, 
we propose an alternative way of treating the above moduli problem in order to remove the 
redundancy which causes the nonseparability of ${\mathscr H}_{\rm diff}$. For this, we introduce 
a {\it local} redundancy, which encodes the concentration of frame vectors (to be defined later on) 
on the tangent spaces $T_pM$ around points of intersections $p$ of smooth loops $\alpha$ in 
$\mathbb{R}^{3}$.

%%%%%%%%%%%%%%%%%%%%%%%%%%%%%%%%%%%%%%%%%%%%%%%%%%%%%%%%%%%%%%%%
\section{Frame theory in a nutshell}
\label{Sec3}
%\hspace*{\parindent}
\parindent3em
%%%%%%%%%%%%%%%%%%%%%%%%%%%%%%%%%%%%%%%%%%%%%%%%%%%%%%%%%%%%%%%%
When working with vector spaces we are always attached to the concept of a basis. 
Indeed, through a basis we can rewrite any vector as a linear combination of 
such vectors of basis. However, the conditions to a basis are very restrictive:
one requires that the elements are linearly independent, and very often we
even want them to be orthogonal with respect to an inner product. This
makes it hard, or even impossible, to find bases satisfying extra conditions,
and this is the reason that one might wish to look for a more flexible tool.
{\it Frames}~\cite{Ch,Mallat} assume this role. In linear algebra, a frame in
a Hilbert space can be seen as a generalization of the idea of a basis to sets 
which may be linearly dependent. A frame allows 
us to represent any vector as a set of ``frame coefficients,'' and to reconstruct a
vector from its coefficients in a numerically stable way. 

Below, following the Ref. \cite{Ch}, we briefly review the basic definitions and results related
to finite frames. Let $\mathscr{H}$ denote an $\ell$-dimensional real or complex Hilbert space.
In this finite-dimensional situation, $\{f_i\}_{i=1}^m$ is called a frame for $\mathscr{H}$,
if it is a -- typically, but not necessarily linearly dependent -- spanning set. 
This definition is equivalent to asking for the existence of
constants $0 < A \leqslant B < \infty$ such that
\begin{equation}
A\parallel f\parallel^{2} 
\leqslant \parallel F(f_{i}) \parallel^2 \leqslant B 
\parallel f\parallel^{2}\,\,, \quad \forall f \in {\mathscr H}\,\,,
\end{equation}
where
\[
\parallel F(f_{i}) \parallel^2=\sum_{i\in I}\mid\langle f,f_{i}\rangle\mid^{2}\,\,.
\]
The numbers $A$ and $B$ are called {\it frame bounds}. They are not unique.
The {\it optimal upper frame bound} is the infimum over all upper frame bounds, 
and the {\it optimal lower frame bound} is the supremum over all lower frame bounds. 
Note that the optimal frame bounds actually are frame bounds. In fact,
\[
A\parallel f\parallel^{2} \leqslant \parallel F(f_{i}) \parallel^2\,\,,
\]
guarantees the uniqueness of frame coefficients, since
\[ 
\parallel F(f_i)-F(f_j) \parallel=0 \Longrightarrow \parallel f_i-f_j \parallel=0\,\,,
\]
while
\[
\parallel F(f_{i}) \parallel^2 \leqslant B 
\parallel f\parallel^{2}\,\,,
\]
guarantees the numerical stability in computing frame coefficients, as long as $B$
is reasonable, since
\[ 
\parallel F(f_i+\delta f_i)-F(f_i) \parallel=
\parallel F(\delta f_i) \parallel
\leqslant \sqrt{B} \parallel \delta f_i \parallel\,\,.
\]

\begin{proposition}[{Christensen~\cite[Proposition 1.1.2]{Ch}}]
Let $\{f_{i}\}_{i=1}^{m}$ be a sequence in $\mathscr{H}$. Then $\{f_{i}\}_{i=1}^{m}$ is a frame 
for a vector space $W :=span \{f_{i}\}_{i=1}^{m}$, where $W \subset \mathscr{H}$ is 
a proper subset of $\mathscr{H}$.
\end{proposition} 

From this proposition follows the

\begin{corollary}[{Christensen~\cite[Corollary 1.1.3]{Ch}}]
A family of elements $\{f_{i}\}_{i=1}^{m}$ in $\mathscr{H}$ is a frame for $\mathscr{H}$ 
if and only if span$\{f_{i}\}_{i=1}^{m}=\mathscr{H}$.
\label{Cor4}
\end{corollary}

According to Corollary \ref{Cor4}, a frame could support more elements 
than is demand for a basis, since the only requirement is that the elements generate
all the space. In particular, given a frame $\{f_{i}\}_{i=1}^{m}$ for $\mathscr{H}$ and 
$\{g_{i}\}_{i=1}^{n}$ an arbitrary collection of vectors in $\mathscr{H}$, then, 
$\{f_{i}\}_{i=1}^{m} \bigcup \{g_{i}\}_{i=1}^{n}$ also is a frame for $\mathscr{H}$. A frame 
that is not a basis is called overcompleted, or {\it redundant}.

Let us consider a Hilbert space $\mathscr{H}$, $\ell$-dimensional, equipped with a frame 
$\{f_{i}\}_{i=1}^{m}$, with $m \geqslant \ell$, and define the following application:
\begin{equation}
T: \mathbb{C}^{m} \rightarrow \mathscr{H}, \quad T\{c_{k}\}_{k=1}^{m} =
 \sum_{k=1}^{m}c_{k}f_{k}\,\,,
\end{equation}
where T is called  {\it pre-frame} operator and its adjoint is given by
\begin{equation}
T^{*}:\mathscr{H} \rightarrow \mathbb{C}^{m}, \quad 
T^{*}f=\bigl\{\langle f,f_{k}\rangle\bigr\}_{k=1}^{m}\,\,.
\end{equation}
 
Notice that in terms of the frame operator, we have
\begin{equation}
\langle Ff,f \rangle=\sum_{k=1}^{m} \mid \langle f,f_{k} \rangle \mid^{2}, 
\quad \forall f \in \mathscr{H}\,\,.
\end{equation}
Thus, the lower frame bound might be sought as a kind of the lower bound of the frame 
operator. A frame $\{f_{k}\}_{k=1}^{m}$ is called tight if we choose $A$ $=$ $B$ in 
the frame definition, that is,
\begin{equation}
\sum_{k=1}^{m}\mid \langle f,f_{k}\rangle \mid^{2}= 
A \|f\|^{2}\,\,.
\end{equation}
Notice that besides the tight requirement, we are assuming the optimal frame bounds
given by the infimum and the supremum. We should also notice that the upper bound condition is 
always assured by the Cauchy-Schwarz-Bunjakowski inequality. Hence, our task is to 
determine the lower frame bound. Thus, we consider the following

\begin{proposition}[{Christensen~\cite[Proposition 1.1.4]{Ch}}]
Let $\{f_{k}\}_{k=1}^{m}$ be a tight frame for $\mathscr{H}$ with frame bound $A$. 
Let $F=A\un$, where $\un$ is the identity operator in $\mathscr{H}$, then
\begin{equation}
f=\frac{1}{A} \sum_{k=1}^{m}\langle f, f_{k}\rangle f_{k}\,\,, 
\quad \forall f \in \mathscr{H}\,\,.
\end{equation}
\end{proposition}
  
We see that the above representation is quite similar to the representation of an 
orthonormal basis with the factor $1/A$ which makes the difference. 

\begin{theorem}[{Christensen~\cite[Theorem 1.1.5]{Ch}}]
Let $\{f_{k}\}_{k=1}^{m}$ be a frame for $\mathscr{H}$ with a frame operator $F$. Then the 
following properties are satisfied:
\begin{itemize}
\item[(i))] $F$ is invertible and self-adjoint.
\item[(ii)] Every $f \in \mathscr{H}$ can be represented as
\end{itemize}
\begin{equation}
f=\sum_{k=1}^{m}\langle f, F^{-1}f_{k} \rangle f_{k} = \sum_{k=1}^{m} \langle 
f, f_{k} \rangle F^{-1}f_{k}\,\,.
\end{equation}
\end{theorem}
  
Every frame in a finite space contain a subset which is a basis. Hence, if $\{f_{k}\}_{k=1}^{m}$ 
is a frame but not a basis, there exists a sequence of nonzero elements $\{d_{k}\}_{k=1}^{m}$ 
such that $\sum_{k=1}^{n}d_{k}f_{k}$ $=$ $0$. So, for all element $f \in \mathscr{H}$ 
we can write it as
\begin{align*}
f&=\sum_{k=1}^{m}\langle f, F^{-1}f_{k} \rangle + \sum_{k=1}^{m}d_{k}f_{k}\\[0.1cm]
&=\sum_{k=1}^{m}\Big{(}\langle f,F^{-1}f_{k} \rangle + d_{k}\Big{)}f_{k}\,\,.
\end{align*}
This shows us that $f$ has many representations as superposition of elements of the frame. 
We shall verify in some examples, quite simple, how to use these frames in order to 
``eliminate'' the redundancy of basis overcompleted and see how these can be applied in 
the case of high valence graphs.

%%%%%%%%%%%%%%%%%%%%%%%%%%%%%%%%%%%%%%%%%%%%%%%%%%%%%%%%%%%%%%%%
\section{\bf The Grot-Rovelli case revisited}
%\section{\bf Preliminares}
\label{Sec4}
%\hspace*{\parindent}
\parindent3em
%%%%%%%%%%%%%%%%%%%%%%%%%%%%%%%%%%%%%%%%%%%%%%%%%%%%%%%%%%%%%%%%
To begin with, let us consider a three-dimensional space with a basis $\{\boldsymbol{v}_k\}_{k=1}^{3}$ 
and let $\bigl\{\boldsymbol{v}_k,\sum_{k=1}^{3}\alpha_{k}\boldsymbol{v}_k,\sum_{k=1}^{3}\beta_{k}\boldsymbol{v}_k\bigr\}$ 
be a frame. From the frame condition we obtain that
\begin{align*}
A\|f\|^{2}& \leqslant \sum_{k=1}^{5} \mid \langle f,f_{k} \rangle \mid^{2}\\[0.1cm]
A\|f\|^{2} & \leqslant \|f\|^{2} + 
\sum_{k=1}^{3}\Big{(}\mid\alpha_{k}\mid^{2} + 
\mid\beta_{k}\mid^{2}\Big{)}\mid\langle f,\boldsymbol{v}_k\rangle\mid^{2}\\[0.1cm]
A & \leqslant 1 + 
\frac{\sum_{k=1}^{3}\Big{(}\mid\alpha_{k}\mid^{2} + 
\mid \beta_{k} \mid^{2} \Big{)}\mid\langle f,\boldsymbol{v}_k \rangle \mid^{2}}
{\|f\|^{2}}\,\,.
\end{align*}
We can see that all redundant information is constrained in the 
frame bounds, and since these bounds are not unique we can take it as optimal ones. 
In this case we should take the infimum of the values
\begin{equation}
\inf\left\{1 + \dfrac{\sum_{k=1}^{3}\Big{(}\mid\alpha_{k}\mid^{2} + 
\mid\beta_{k}\mid^{2}\Big{)}\mid \langle f,\boldsymbol{v}_k\rangle \mid^{2}}
{\|f\|^{2}}\right\}\,\,,
\end{equation}
where we can eliminate the ``extra'' information because in this case $A$ $=$ $1$. 
This process can be done for $n$ redundant elements and it is interesting that we can 
restore from the frame bounds only the necessary ``information''. We can also notice that 
we had, initially, a overcomplete set from the perspective of basis. Nonetheless, when 
looking at it as a frame and applying the definitions we recover the expansion only in 
terms of the basis elements. Moreover, we could add to a basis any number of elements, 
linearly dependent, which in the frame point of view we recover only what is really necessary, 
in other words (as in signal analysis) we can filter all redundant information that makes 
the set overcomplete.
   
Therefore, if we consider a $Q$-valence graph in a three-dimensional
hypersurface, we could look at the tangent vectors at the knot as forming a frame and as shown above,
filter the information which generates the moduli structure and the nonseparability of 
${\mathscr H}_{\rm diff}$, without need to extend the diffeomorphism group.

%%%%%%%%%%%%%%%%%%%%%%%%%%%%%%%%%%%%%%%%%%%%%%%%%%%%%%%%%%%%%%
\subsection{Redundancy}
%\hspace\parindent
\parindent3em
%%%%%%%%%%%%%%%%%%%%%%%%%%%%%%%%%%%%%%%%%%%%%%%%%%%%%%%%%%%%%%
The redundancy of a frame, at least in finite dimensions, is a precise quantitative notion.
In fact, the following theorem shows that the redundancy of a frame can be measured through 
the frame bounds $A$ and $B$.

\begin{theorem}[{Mallat~\cite[Theorem 5.2]{Mallat}}]
In a space of finite dimension $N$, a frame of $M \geqslant N$ normalized vectors has 
frame bounds $A$ and $B$, which satisfy
\begin{equation}
A\leqslant \frac{M}{N}\leqslant B\,\,.
\end{equation}
\label{Theo7}
\end{theorem}
For a tight frame $A=B=M/N$. As this theorem is valid for any 
finite $M$, we can apply it in the situation treated in Ref.~\cite{FR} in order 
to avoid moduli structures, without requiring the extension of the diffeomorphism group 
at a quantum level.

%%%%%%%%%%%%%%%%%%%%%%%%%%%%%%%%%%%%%%%%%%%%%%%%%%%%%%%%%%%%%%
\subsection{$Q$-Valence Graphs in LQG}
%\hspace\parindent
\parindent3em
%%%%%%%%%%%%%%%%%%%%%%%%%%%%%%%%%%%%%%%%%%%%%%%%%%%%%%%%%%%%%%
Now consider a $Q=4$ valence graph in LQG where three of them are linearly independent,
for example, $\boldsymbol{v}_1,\boldsymbol{v}_2,\boldsymbol{v}_3$. We 
shall denote them as $\{\varphi_{i}\}=\bigl\{\boldsymbol{v}_1,\boldsymbol{v}_2,\boldsymbol{v}_3,
\boldsymbol{v}_4=\sum_{i=1}^3\alpha_{i}\boldsymbol{v}_i\bigr\}$. Then, 
for any vector $f$ on $\mathscr{H}$ we have
\begin{align*}
\sum_{i=1}^{4} \mid \langle f,\varphi_{i}\rangle\mid^{2} &= 
\sum_{i=1}^{3}\mid\langle f,\boldsymbol{v}_i\rangle\mid^{2} + 
\mid\langle f,\sum_{i=1}^{3} 
\alpha_{i}\boldsymbol{v}_i\rangle\mid^{2}  \\[0.1cm]
& = \|f\|^{2} + \mid \sum_{i=1}^{3}\langle f, \alpha_{i}\boldsymbol{v}_i\rangle\mid^{2} \\[0.1cm]
& \leqslant \|f\|^{2} + 
\sum_{i=1}^{3}\alpha_{i}^{2}\mid\langle f, \boldsymbol{v}_i\rangle\mid^{2} \\[0.1cm]
& \leqslant \sum_{i=1}^{3} (1+ \alpha_{i}^{2})\mid\langle f, \boldsymbol{v}_i\rangle\mid^{2}\,\,.
\end{align*}
On the other hand, by the Theorem \ref{Theo7}, if we are dealing with a tight frame, then
\begin{equation}
\sum_{i=1}^{4} \mid \langle f,\varphi_{i}\rangle\mid^{2} = 
\frac{4}{3}\sum_{i=1}^{3}\mid\langle f, \boldsymbol{v}_i\rangle\mid^{2}\,\,.
\end{equation}
Note that the expansion coefficients of the redundant vector 
yield the following value for $\alpha$:
\begin{equation*}
\alpha_{i}^{2} = \frac{1}{3}\,\,.
\end{equation*}

Finally, let us consider again the simple case
considered by Grot-Rovelli~\cite{GR} in which a moduli space appear,
which is a five valence graph. Hence, we have an intersection point $p$ 
crossed by the loop $\alpha$ five times. Let us fix an arbitrary coordinate chart 
in the neighborhood of $p$, and let $\boldsymbol{v}_i$ a for $i=1,2,3,4,5$ be
the components of the five tangents of the loop $\alpha$ at $p$. 
Assume any three of the five vectors are linearly independent, again
$\boldsymbol{v}_1,\boldsymbol{v}_2,\boldsymbol{v}_3$ (these three vectors 
define a basis in the tangent space at $p$), then the set
$\{\varphi_{i}\}=\bigl\{\boldsymbol{v}_1,\boldsymbol{v}_2,\boldsymbol{v}_3,
\boldsymbol{v}_4=\sum_{i=1}^3\alpha_{i}\boldsymbol{v}_i,
\boldsymbol{v}_5=\sum_{i=1}^3\beta_{i}\boldsymbol{v}_i\bigr\}$ is used to build 
the frame. Thus, we obtain a constrained redundancy
\begin{equation*}
\alpha_{i}^{2} + \beta_{i}^{2} = \frac{2}{3}\,\,.
\end{equation*}

Naturally, for an $Q$-valence graph in LQG (with a finite set of 
redundant vectors) we obtain that
\begin{equation*}
\sum_{j=1}^{M-N}\alpha_{i,j}^{2} = \frac{(M-N)}{N}\,\,,
\end{equation*}
where $\alpha_{i,j}$ is the $i$-th coefficient of the $j$-th redundant vector,
$M$ is the total number of elements in the frame (basis + redundancy) and $N$ is 
the dimension of the space. Thus, we see 
that through the perspective of a frame we expand a set, \textit{a priori}, larger than 
a basis in a stable way and convert the redundancy into a {\it multiplicative coefficient} 
that measures and controls the problem of high valence graphs in LQG. 
In turn, since the redundant vectors can be removed in a stable way, we do not have more the 
problem of the mapping between tangent spaces that requires linearity. In fact, the 
redundancy {\it does not change} once an invertible operator is applied to a frame~\cite{BCK}.
Thus, the tangent map $f^{*}:T_pM \rightarrow T_qM$ is allowed and we do not need to extend 
the diffeomorphism group. In conclusion, by using the setting of frames, the
moduli structures are not anymore present, and the kinematical Hilbert space of the 
diffeomorphism invariant states of LQG is separable.

%%%%%%%%%%%%%%%%%%%%%%%%%%%%%%%%%%%%%%%%%%%%%%%%%%%%%%%%%%%%%%%%%%%%%%%%%%%

%%%%%%%%%%%%%%%%%%%%%%%%%%%%%%%%%%%%%%%%%%%%%%%%%%%%%%%%%%%%%%%%%%%%%%%%%%%%%%

%%%%%%%%%%%%%%%%%%%%%%%%%%%%%%%%%%%%%%%%%%%%%%%%%%%%%%%%%%%%%%%%%%%%%%%%%%%%%%
\end{document}